\documentclass[a4paper,11pt]{article}

\usepackage{jinstpub}
\usepackage{graphicx}
\usepackage{xcolor}
\usepackage{caption}
\usepackage{subcaption}
\usepackage{multirow}
\usepackage{verbatim}
\usepackage{textgreek}

\author[a,b]{Li Wang,}
\author[b,1]{Jilei Xu,\note{Corresponding author}}
\author[a,2]{Shuxiang Lu,\note{Corresponding author}}
\author[b]{Haoqi Lu,}
\author[b]{Zhimin Wang,}
\author[b,c,3]{Min Li,\note{Now at IPHC, Universit\'{e} de Strasbourg, CNRS/IN2P3, F-67037 Strasbourg, France}}
\author[b]{Sibo Wang,}
\author[b,c]{Changgen Yang,}
\author[b,c]{Yichen Zheng}
\affiliation[a]{Zhengzhou University, zhengzhou, 450001, China}
\affiliation[b]{Institute of High Energy Physics, Beijing, 100049, China}
\affiliation[c]{University of Chinese Academy of Sciences, Beijing, 100049, China}
\emailAdd{xujl@ihep.ac.cn,lushuxiang@zzu.edu.cn}

\title{A novel design for 100 meter-scale water attenuation length measurement and monitoring}
\abstract{Water Cherenov detector is a vital part in most of neutrino or cosmic ray research. As detectors grow in size, the water attenuation length (WAL) becomes increasingly essential for detector performance. It is essential to measure or monitor the WAL. While many experiments have measured WAL in the lab or detector, only the Super-Kamiokande experiment has achieved values exceeding 50 meters in the detector with a moving light source. However, it is impractical for many experiments to place a moving light source inside the detector, necessitating an alternative method for investigating long WAL. A novel system has been proposed to address the challenge of investigating long WAL. This system focuses on ample water Cherenkov detectors and features a fixed light source and photomultiplier tubes (PMTs) at varying distances, eliminating the need for moving parts. The static setup demands high precision for accurate measurement of long WAL. Each component, including LED, diffuse ball, PMTs, and fibers, is introduced to explain uncertainty control. Based on lab tests, the system's uncertainty has been controlled within 5\%.
Additionally, camera technology is also used during the evaluation of the system uncertainty, which has the potential to replace PMTs in the future for this measurement. Monte Carlo simulations have shown that the system can achieve a 5\% measurement uncertainty at WAL of 80 meters and 8\% at WAL of 100 meters. This system can be used in experiments with large Cherenkov detectors such as JUNO water veto and Hyper-K.}

\keywords{Water attenuation length; Water Cherenkov; diffuse ball; non-uniformity}

\begin{document}

\maketitle

\flushbottom

\section{Introduction}\label{sec1}

In high-energy physics experiments, the water Cherenkov detector plays a crucial role. With the development of detecting reactor neutrino, astroneutrino and primary cosmic ray, the water Cherenkov detector is becoming larger and larger, the water attenuation length (WAL, also known as water transparency) is the most critical parameter for monitoring and verifying the quality of water. In previous experiments for measuring water or liquid scintillator (LS) attenuation length, one kind of experimental setup usually is built in the lab, and the other is built directly in the liquid detector. 

For the setup in the lab, the general method is using an LED and a focus lens to generate parallel light that passes through a vertical tube filled with liquid at different depths, such as the LS attenuation length measured in Daya Bay (DB) \cite{LiuJinchang} and in JUNO \cite{YuBoxiang-PMTAttL, HYang-junoLS-JINST, XiangweiYin-junoLS-RDTM}, the WAL measured in LHAASO \cite{WCDA, Anqi} (Method 1 in Tab.~\ref{tab:experiments}). Usually, the light is detected by a PMT at the bottom and the attenuation lengths are measured at a few tens of meters. There also has a WAL measurement device in CHIPS \cite{chips}, which uses a vertical tube with 3 meters, a laser as a light source, a photodiode as a photosensor, and the WAL can be measured up to 139 meters, but the uncertainty is large, $\pm$ 42 m (Method 2 in Tab.~\ref{tab:experiments}). These vertical tube devices are usually put in the lab, just to test the water sample. To get the more precise WAL, a longer vertical tube is needed. But it will be harder to increase the height of the vertical tube and control the device to measure a longer attenuation length, like 80~m - 100~m long scale. In recent years, a WAL test device has been developed with a horizontally placed water container (length of 8~m) in the experiment of LHAASO \cite{LHAASO}, which had multiple light sources with a collimator to make the light parallel. The device measured WAL in 20 meters with an uncertainty of about 3\%, and the deduced WAL 100 m with an uncertainty of 20\% (Method 3 in Tab.~\ref{tab:experiments}). For the long WAL measuring, as long as the ultrapure water is sampled from the detector, the WAL will be quickly decreased in the container or vertical tube. It is not easy to get the real WAL of the detector in time.  

For the setup in the detector, only the Super-Kamiokande (Super-K) experiment \cite{Super-KDet} gave a precise measurement. It used an LED plus diffuse ball (also known as a diffuser) as the light source and a camera as the photosensor. It depends on the diffuser moving up and down to get the different light propagation distances. The WAL reaches nearly 100 meters, 97.9 $\pm$ 3.5 m, which shows excellent ultrapure water circulation and operation (Method 4 in Tab.~\ref{tab:experiments}). Besides in ultrapure water, the WAL is required to be measured in a deep sea, for example in TRIDENT \cite{Hailing} and ANTARES \cite{ANTARES} experiments. The diffuser and photosensors are immersed in the sea and pulled by a long cable. The WAL of the sea is generally about 20 - 40 meters, with a maximum of 60 meters (Method 5 in Tab.~\ref{tab:experiments}). Also, a method for measuring the attenuation length of liquid material using a camera has been developed \cite{Super-KDet, Hailing, YuBoxiang-LS}. The camera can replace the PMT to be the photosensor. 

These measurements are summarized in Table \ref{tab:experiments}. The light source has two different styles of lenses with parallel light or diffuse balls with point-like (spherical isotropic) light. The container is a vertical tube, horizontal box or the direct Cherenkov detector. For the vertical tube devices in the lab, the lens needs to be adjusted to focus the light and make the light spot glow in the sensitive area of the photosensor when changing the water depth. It is hard to increase the height of the tube (> 8~m) to give a more precise measurement of long WAL. The device in the lab can't test the long WAL in time either. For the devices in the Cherenkov detector, the moving operation of the light source or photosensors is not easy to operate to monitor the water quality online all the time. The moving facility should make sure always waterproofing and working. The mechanical structure is complex and difficult to maintain. In this paper, we proposed a novel device that can be directly fixed in the water Cherenkov detector to give online monitoring and does not need the moving operation of both light source and photosensors (Method 6 in Tab.~\ref{tab:experiments}).  The detail of the proposal is introduced in section \ref{sec2}; the device R\&D is introduced in section \ref{sec-R&D}; the system performance anticipation is introduced in section \ref{sec-anticipation}; and section \ref{sec7} is the conclusions.

	\begin{table}[!htb]
	\centering
	\caption{Summary of the water or LS attenuation length measurements.}
	\label{tab:experiments}
	 \vspace{0.2cm}
	\begin{center}
	\resizebox{\textwidth}{!}
{
	\begin{tabular}{ccccccccc}
			\hline
			\hline
		Method & Light Source & Photosensor & Container & Work for &Work in &Characteristic& WAL (m)& Exp. (e.g.) \\
		\hline
        1 & LED + Lens &PMT & Vert. tube & Water/LS & Lab &Change water depth& 0 - 30& DB\cite{LiuJinchang}, LHAASO \cite{WCDA}\\
        &&or Camera&&&&&& JUNO \cite{YuBoxiang-PMTAttL, HYang-junoLS-JINST, XiangweiYin-junoLS-RDTM, YuBoxiang-LS}\\
        2 & Laser + Lens &Photodiode & Vert. tube & Water & Lab &Change water depth&139 $\pm$ 42& CHIPS \cite{chips}\\

        3 & LEDs + Clmt. & PMT & Hori. box & Water & Lab &Multi light source&17 $\pm$ 0.5& LHAASO \cite{LHAASO}\\  
        % &  different places &  &  &   &  & \\ 
        4 & Laser + Diffuser&Camera & Vert. in pool & Water & Detector &Move diffuser&97.9 $\pm$ 3.5& Super-K \cite{Super-KDet}\\
        5 & LED + Diffuser & 2 Cameras & Vert. in sea & Water & Detector &Move PMTs&20 - 60& TRIDENT \cite{Hailing}\\
         &&or PMTs&&&&&& ANTARES \cite{ANTARES}\\
        6 & LED + Diffuser & 8 PMTs & Hori. in pool & Water& Detector &Static&& Proposed here\\  
        &  & or cameras  & &  && \\       
        \hline
                \hline
	\end{tabular}
	}
\end{center}
	\end{table}

\section{A proposal of WAL measurement and monitor in ultrapure water}\label{sec2}

\begin{figure}
  \centering
  \includegraphics[width=0.6\textwidth]{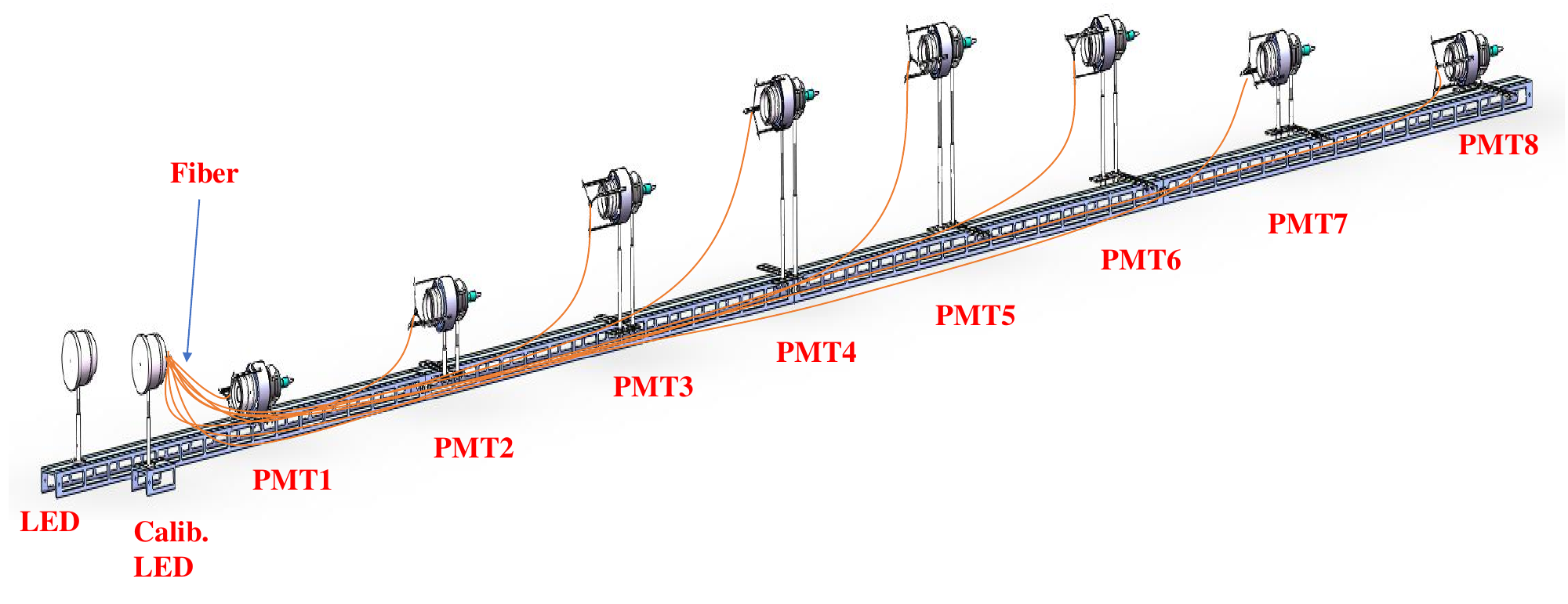}
  \hspace{0.2in}
  \includegraphics[width=0.3\textwidth]{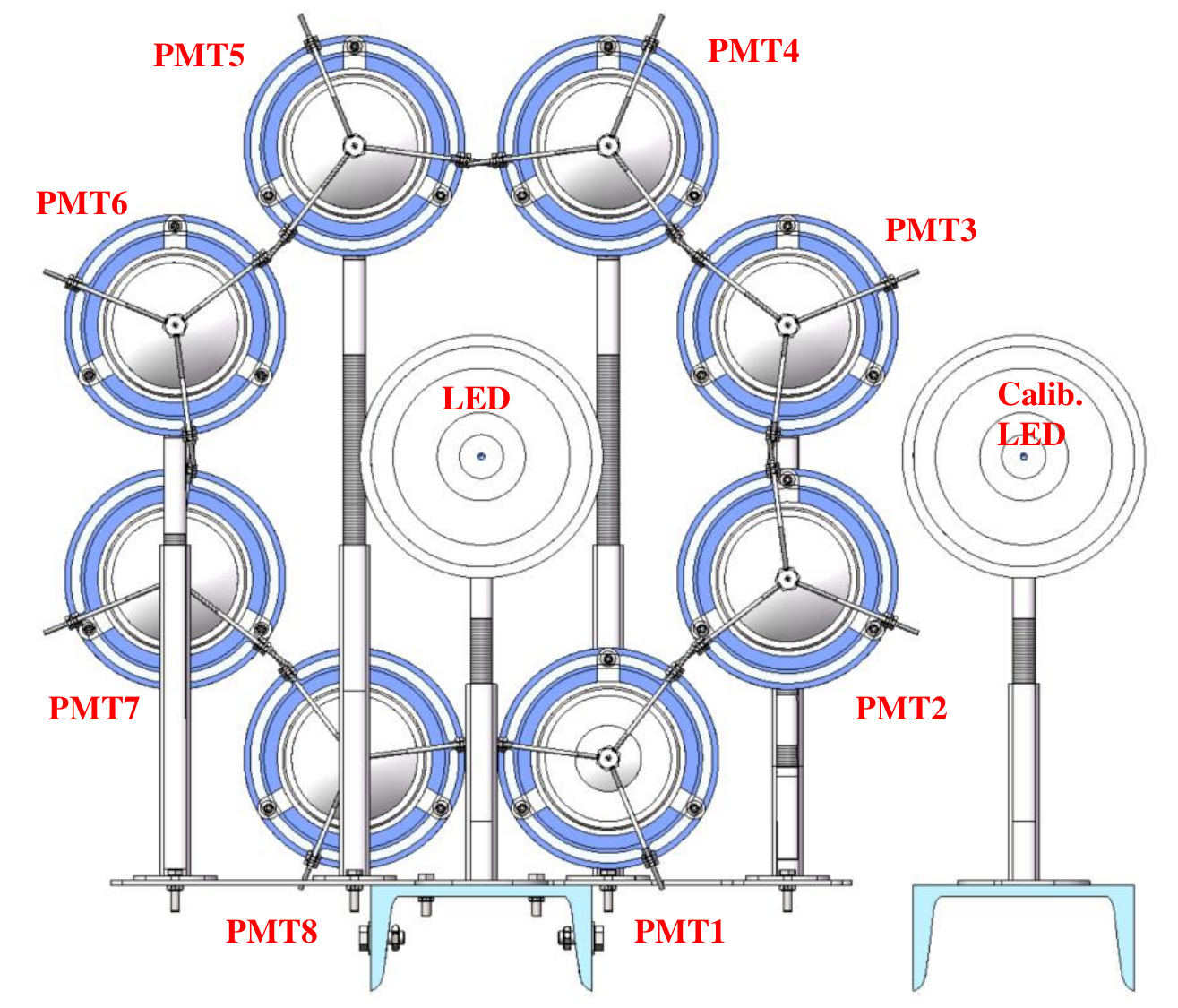}
  \caption{The proposed device of measurement.}
  \label{fig:proposal}
\end{figure}

We propose for the first time a static device for measuring the WAL using a light-diffusing ball and several PMTs (8 PMTs here) placed at varying distances (Fig.~\ref{fig:proposal}). The LED serves as the light source at one end, while the PMTs are positioned at different distances from the LED. To ensure that all PMTs receive the LED pulse light simultaneously, they are placed along a circle, rather than in a straight line, when viewed from the light source. The uniformity of the light source is essential for the success of the measurement, and a point-like source is required because the circle radius exceeds the size of the light source. Therefore, a diffuse ball is necessary to cover the LED and produce more uniform light. This ensures that if the group of circled PMTs are placed at the same distance, they will all be covered by the same light density. The non-uniformity of the diffuse ball is expected to be better than 2\%.

For the PMTs at varying distances, if the device is in air, the light density will decrease according to the inverse-square law, $d^{-2}$, where $d$ represents the distance from the LED to the PMT. This can be utilized to validate the system error. When the device is submerged in water, the light also follows an exponential decay function, resulting in the overall function obeying the equation:
\begin{equation}
    Q(d)=Ad^{-2}e^{-{d}/{\lambda}},
    \label{eq:attL}
\end{equation}
where $Q$ is the charge collected by the PMT; $\lambda$ is the WAL; $A$ is the charge coefficient. For the real data taking, each PMT has its own $d$ and gets its $Q$. Then we can use this function to fit out the WAL, $\lambda$. 

Since the different PMTs have different performances, such as quantum efficiency, collection efficiency, gain, and so on, they need to be calibrated in advance. But, for a long time working in the water, PMT performance will change with time like aging. We proposed the other LED be used to calibrate. The calibration LED was packed and connected to a group of fibers. The photons are just allowed to go into the fiber, and then arrive at each PMT cathode through each fiber. The fibers have the same length to get the relatively same photon number when light arrives at the PMT. Although the fibers have the same length, the fiber will be calibrated by a PMT to get the fiber differences. The PMTs and their electronics also will be calibrated by a fiber to get the PMT plus electronics differences. In the formal data analysis, the difference will be eliminated by the calibration data. 

To measure longer attenuation lengths requires that the measurement device is also longer. We propose the total length can be 30~m, which can be used in the large water Cherenkov experiment, such as Hyper-K, and JUNO.  The whole system has been simulated with the toy Mont Carlo (toyMC). When the total system uncertainty is within 5\%, the WAL is the order of a hundred meters, the measurement uncertainty can be 8\% (see section~\ref{sec-anticipation}).

In recent years, with the development of the camera, it has become more and more popular and has a potential to instead of PMT for attenuation length measurements (Experiments with cameras in table \ref{tab:experiments}). We introduce it in this paper, also because it is used in our testing to verify the optical testing comparison with the PMTs. The study and understanding of the camera performance provide an important experience for the camera application to WAL measurement. 

\section{The WAL device R\&D} 
\label{sec-R&D}

\subsection{Diffuser uniformity test}
\label{sec-uniformityTest}

%\subsection{Camera linearity study}
%\label{Principle-camera}
The camera imaging technology was used to study the diffuser uniformity for convenient operation. A CCD camera (brand WORK POWER) \cite{CCDfactory} was selected in our laboratory since its larger size of photosensor and the low noise than Complementary Metal-Oxide-Semiconductor Transistor (CMOS)  \cite{CCD photoreceptor area, CMOS and CCD}. The characteristics of different cameras are different, such as the phenomenon of phototropic supersaturation \cite{CCD Excessive saturation}, the light intensity response characteristics, etc. First, they need to be studied to reduce the measurement error. 

\begin{figure}[!hbt]
  \centering
  \includegraphics[width=0.8\textwidth]{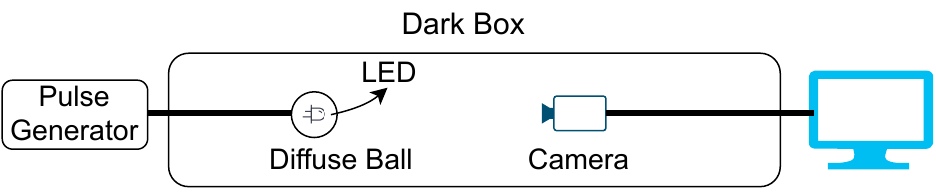}
  \caption{Diagram of the CCD imaging system. The diffuser and the camera are put in the dark box.}
  \label{fig:setup}
\end{figure}

\begin{figure}[!hbt]
  \centering
  \includegraphics[width=0.5\textwidth]{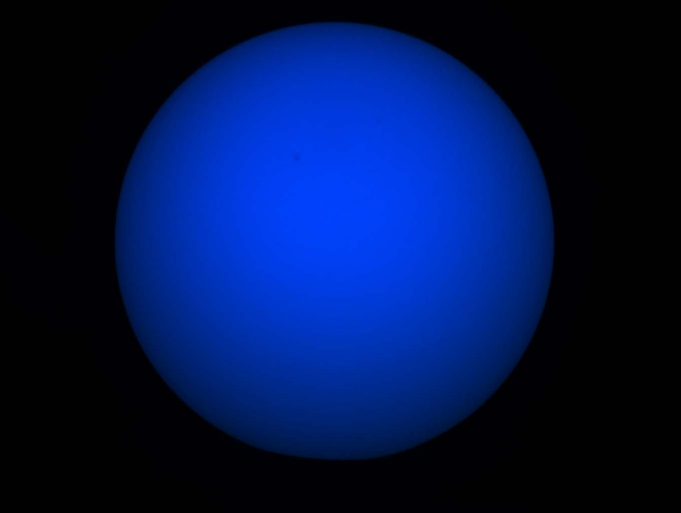}
  \caption{The image of a light source by the CCD camera. The total pixel size is 1360 $\times$ 1024. Exposed time is 0.2~s.  }
  \label{fig:DiffuserImage}
\end{figure}

We build a camera system to image the diffuse ball when the LED is turned on (Fig.~\ref{fig:setup}). The LED covered with the diffuse ball is put at one end of the dark box. The camera is put on the other side facing the diffuser. The camera lenses need to be adjusted in focus before imaging. The diffuser image is blue color because the LED light wavelength is about 400~nm (Fig.~\ref{fig:DiffuserImage}).

\begin{figure}
  \centering
  \includegraphics[width=0.5\textwidth]{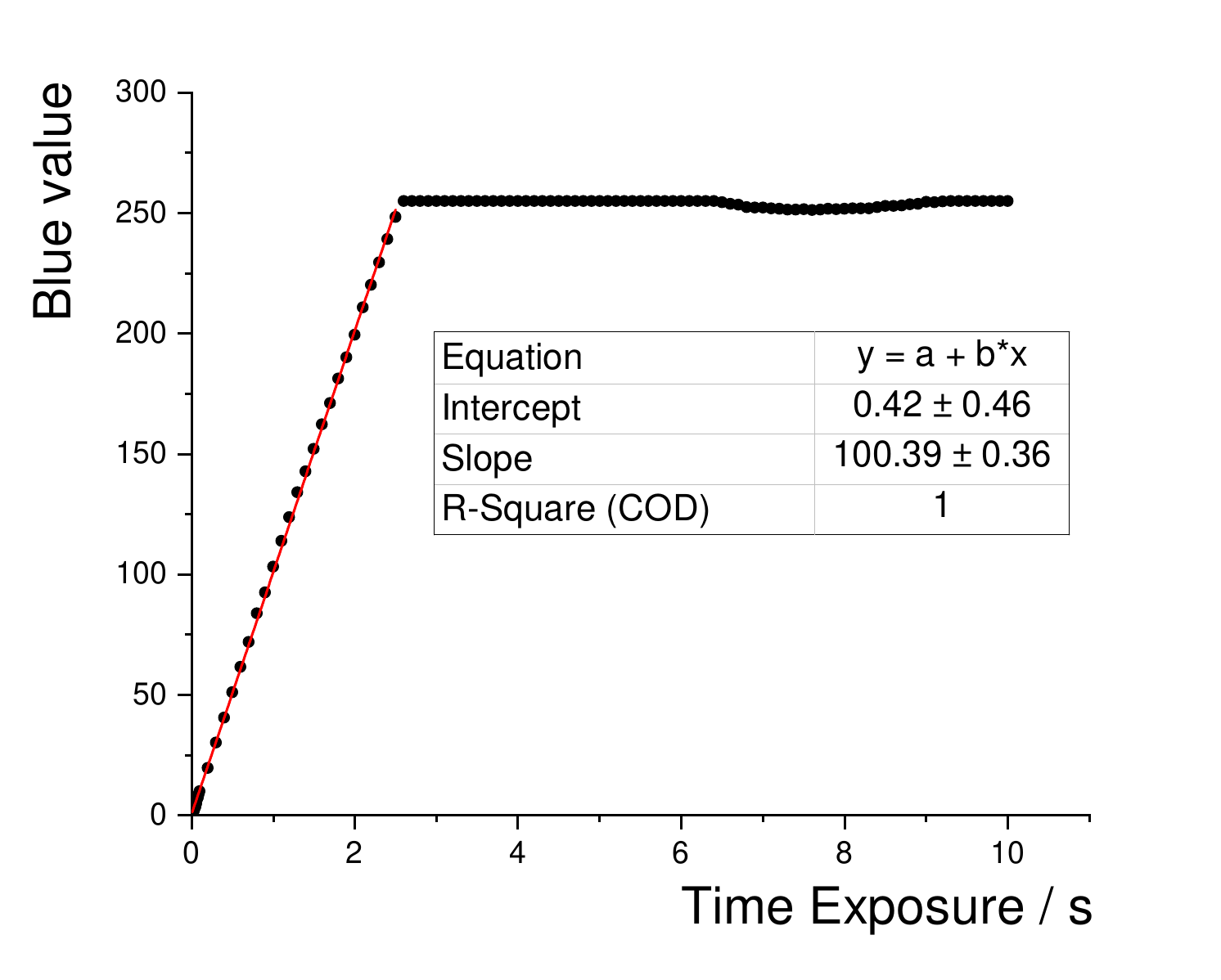}
  \caption{The blue light intensity versus exposure time. The camera working linearity range is from 0.1~s to 2.5~s, fitted with linearity. }
  \label{fig:cameraLinearFit}
\end{figure}

Due to the camera pixel being small, its charge potential well is limited, and it is easy to reach saturation and affect the charge linearity of the camera \cite{Charge-Coupled Devices}. For a color camera, the light intensity values of three channels R, G, B (Red, Green, Blue) are independent of each other \cite{Measuring Colour}. The blue value is extracted in the image study. We exposed the different times from 1~ms to 10~s in the camera linearity study. When the blue channel is saturated, the maximum digitized value is 256. From Fig.~\ref{fig:cameraLinearFit} we can see, in the time range 0.1~s to 2.5~s, the camera has a good linearity because the slope error is smaller than 0.5\%; after 2.5~s, the blue channel is saturated since the charge leakage \cite{CCD charge leakage}. Although the camera tries to compensate for the overexposure in 6~s to 9~s it still can't be used. In the following studies, the camera has been operating in the linearity region all the time.

%\subsection{Diffuse ball and LED uniformity study}

The photons of the camera-exposed image come from the LED light source, which are emitted from the LED, and undergo refraction, scattering, and reflection multiple times in the diffuse ball and then emit at the surface of the diffuse ball. The uniformity of the diffuser is affected by the position of the LED, the emission angle, and the materials of the ball. However, the optical properties of the diffuse ball remain unchanged \cite{Reflectance Spectroscopy}.

The ball is three-dimensional, but the image is two-dimensional. The small area in the image center with a radius of 1~cm (more like a point-like light source), can be approximated as the flat area compared with the ball radius of 4~cm. The image is obtained as Fig.~\ref{fig:setup} and is processed to obtain the value of non-uniformity by the equation:
\begin{equation}
    \delta=(\sigma/M)\times100 \% 
    \label{eq:sigma}
    \end{equation}
    
    Where $\sigma$ is the standard deviation and $M$ is the average light intensity of pixels in the selected area. 
    
\begin{figure}
  \centering
  \includegraphics[width=0.4\textwidth]{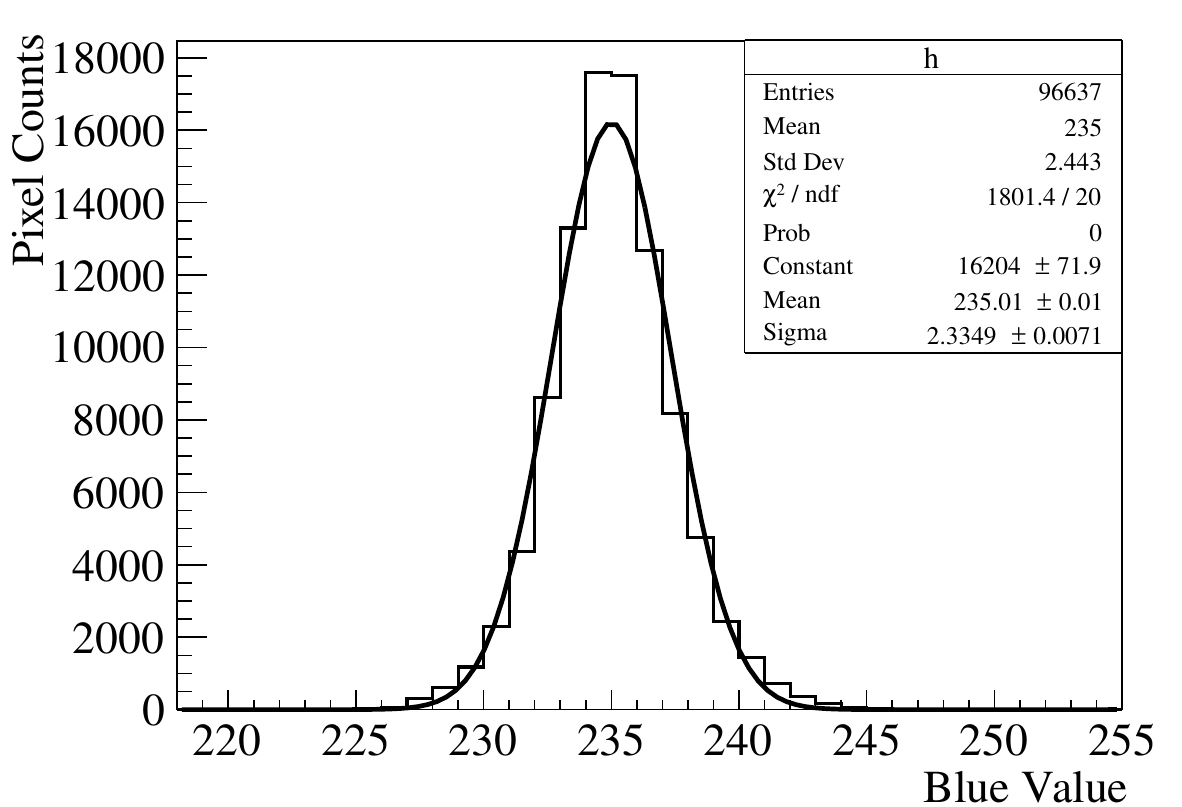}
  \includegraphics[width=0.4\textwidth]{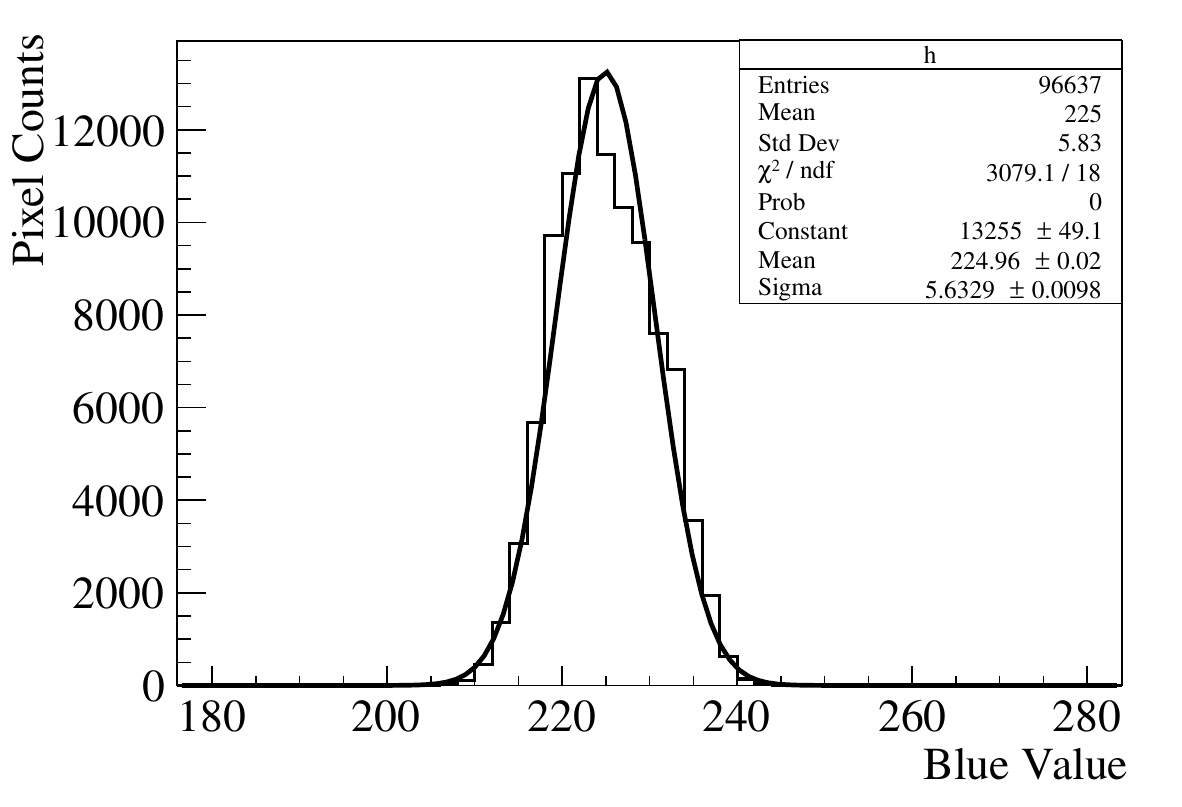}
  \caption{Statistics results of diffuser blue values lighted by straw-hat LED (left) and lamp-bead LED (right), and fit with the Gaussian function. }
  \label{fig:uniformity Fit}
\end{figure}

In the laboratory, two different blue LEDs, straw-hat type and lamp-bead type, are selected from the electronic market \cite{taobao}. The lamp-bead LED luminous angle is 60 degrees smaller than the straw-hat's 120 degrees, but the power is higher than the straw-hat LED. The two LEDs are placed in the same diffuse ball with material of Nylon, respectively. The LED light uniformity was measured by the camera (Fig.~\ref{fig:DiffuserImage}) and the image was analyzed by equation~(\ref{eq:sigma}). The results are shown in Fig.~\ref{fig:uniformity Fit}. The pixel data histogram was fitted by the Gaussian function. The non-uniformity is 0.99\% for straw-hat LED and 2.5\% for lamp-bead LED. The straw-hat LED has smaller non-uniformity and was finally selected.

 Six different diffuse balls are tested. They are two types of Nylon material and polytetra fluoroethylene (PTFE) material with four different sizes or LED positions (Tab.~\ref{tab:LED}). The three balls with a diameter of 80~mm are Nylon with serial 66, 1010 and PTFE respectively. The three PTFE balls with a diameter of 63~mm are dug the different hole depths to put the LED. The same straw-hat LED was used when testing the different diffusers. (Fig.~\ref{fig:DiffuserImage}).

\begin{table}
	\centering
		\caption{Diffuser non-uniformity results for different materials and different LED positions in the diffuse ball. The `R' in the third column means the ball radius and the LED is put in the center of the ball.}
	\label{tab:LED}
	\begin{tabular}{ccccc}
		\hline
		\hline
		Class & Diffuser Dia.& LED position & Non-uniformity \\
				&(mm)  &(mm)  &  (\%) \\
		\hline
        Nylon66 & 80  & R & 0.99 \\
        Nylon1010 & 80  & R & 0.95 \\
        PTFE-1 & 80 &  R & 3.40 \\
        PTFE-2 & 63 &  1/3R & 2.39 \\
        PTFE-3 & 63 &  2/3R & 6.48\\
        PTFE-4 & 63 & R & 3.77 \\
        \hline
        \hline
	\end{tabular}

	\end{table}

  Compared with PTFE-2,3,4, the LED position at `1/3R' is the best, but the image is distorted due to the excessive light on the side of the ball, indicating that the hole is too near the surface of the ball. Except for the `1/3R' configuration, when the LED is in the center position (R), the non-uniformity is smaller than the `2/3R' position. When comparing different materials with the same diameter size of 80~mm, the non-uniformity of Nylon1010 and Nylon66 is at the same level, smaller than 2\%, and much smaller than PTFE. Given that Nylon66 is the more common material, it was ultimately selected.

\subsection{System uncertainty estimation}\label{sec-uncertainty3percent}

The uncertainty of the proposed system is the key to the experiment's success. In the lab, a dark box is built; eight 3-inch PMTs are purchased from HZC Photonis \cite{3-inch-PMT}; 8 fibers with each length 8~m are from XINRUI \cite{fiberFactory}; the PMT data acquisition is by Flash ADC \cite{haiqiong-paper}. Besides the LED and the diffuse ball selected above, each part is integrated to verify the whole system's uncertainty and stability. The testing system is just like the setup in Fig.~\ref{fig:setup}. The difference is the camera is replaced by PMT.

\begin{figure}
  \centering
  \includegraphics[width=0.4\textwidth]{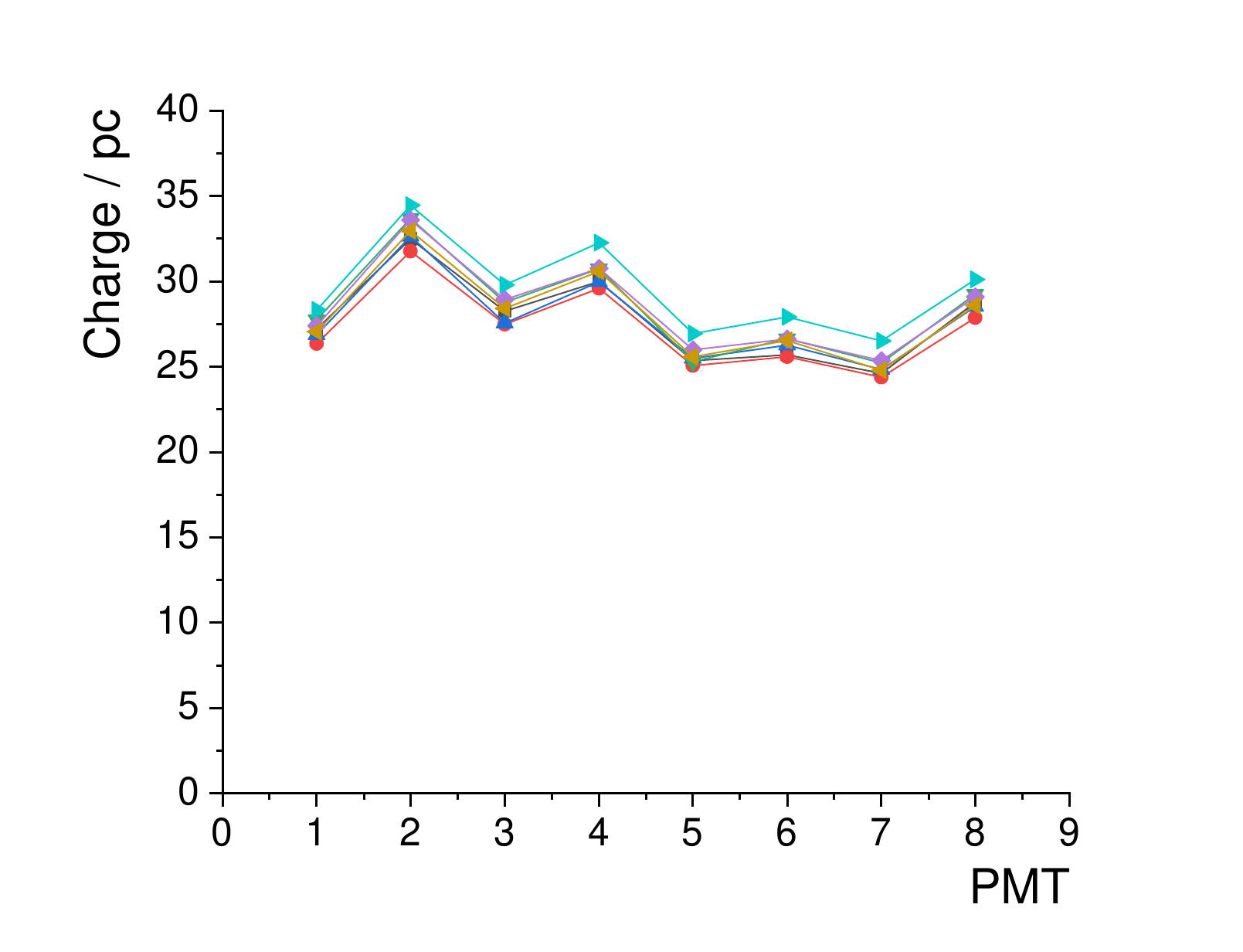}
  \includegraphics[width=0.4\textwidth]{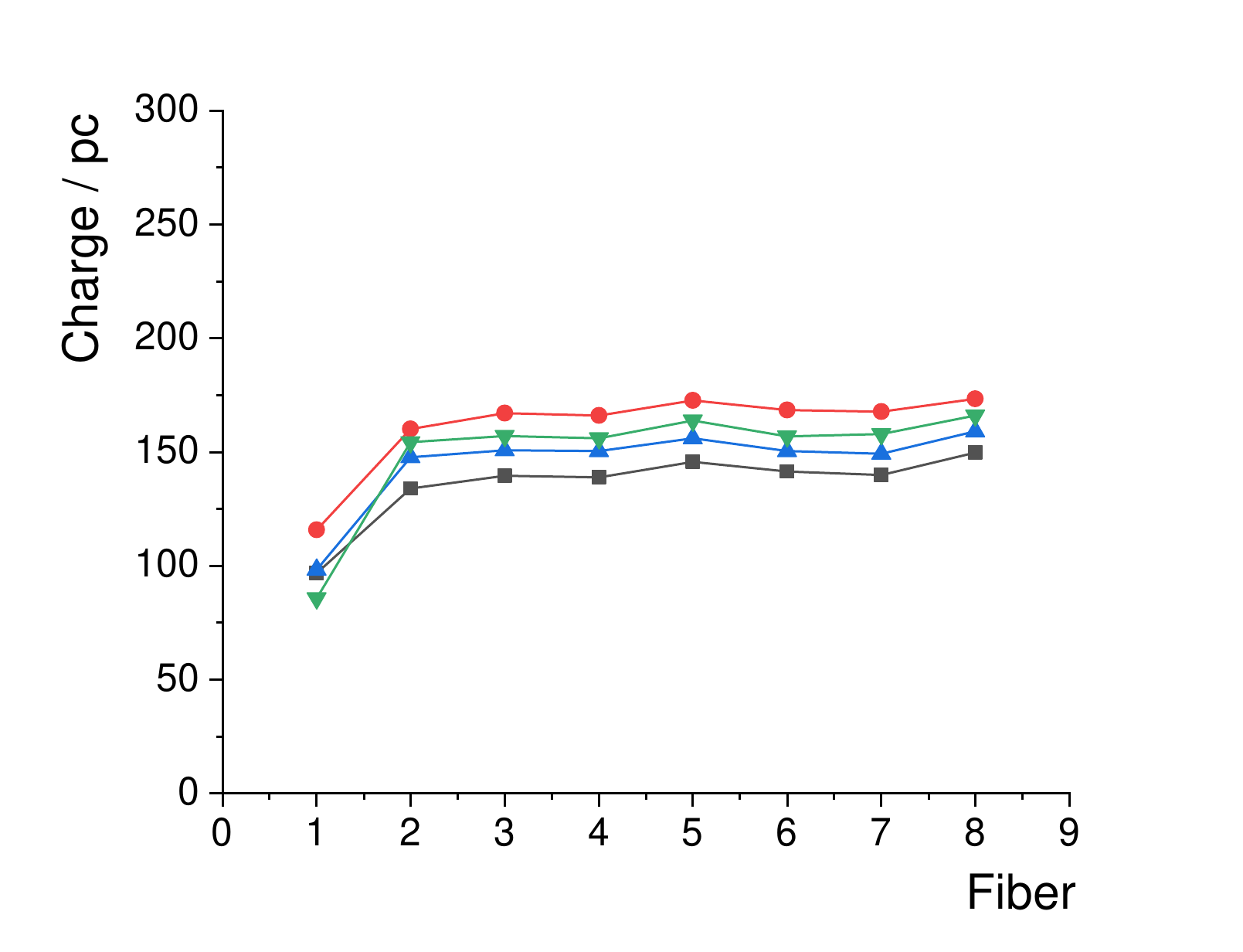}
  \caption{Eight channels of PMT and electronics are calibrated by 7 fibers (left). Each colored line expresses one fiber. Eight fibers are calibrated by 4 PMTs (right). Each colored line expresses one PMT. }
  \label{fig:sub-scales}
\end{figure}

The PMTs are calibrated by LED and adjusted high voltage to let the PMTs work at the gain 3 $\times$ 10$^{6}$. To ensure that all 8 channels are identical, we need to calibrate the system. The system is divided into two parts. One is fibers, and the other is PMTs and electronics. The 8 fibers were tested one by one with the same PMT and electronic channel. They are tested in total 4 times by 4 PMTs (Fig.~\ref{fig:sub-scales}, right). We observe that the colored lines are nearly parallel, indicating the stability of the test. The variation in charge suggests non-uniformity in the fiber, but this can be rectified using the calibrated data. Additionally, the PMTs and electronics are individually tested with the same fiber, with a total of 7 tests using 7 fibers (Fig.~\ref{fig:sub-scales}, left). The parallel colored curves further demonstrate the stability of the test. The charge fluctuation across different channels encompasses differences in PMT detection efficiency, gain, and electronic channel, all of which can be adjusted using the calibrated data for subsequent corrections.

During the calibration process, the charge measured for each PMT over the 7 tests can be averaged, providing the averaged data for each PMT. Subsequently, the data from the eight PMTs can be normalized to the total average, enabling the calculation of each PMT's relative difference ($d_{p}$). Similarly, each fiber's relative difference ($d_{f}$) can be obtained by averaging the 4 times measurements. After each fiber is matched with its own PMT, the LED is lit by a pulse generator, and photons go through 8 fibers to fire the PMTs, then all 8 PMTs can get the signal at the same time. For each PMT, more than 20,000 waveforms with multi-photoelectron (multi-PE)  are collected. After integrating the waveform charge, we can get the average charge of each PMT channel ($Q_{a}$). The difference of each channel can be corrected by the equation:

\begin{equation}
    Q=Q_a/d_p/d_f
    \label{eq:corr}
\end{equation}

\begin{figure}[!hbt]
  \centering
  \includegraphics[width=0.6\textwidth]{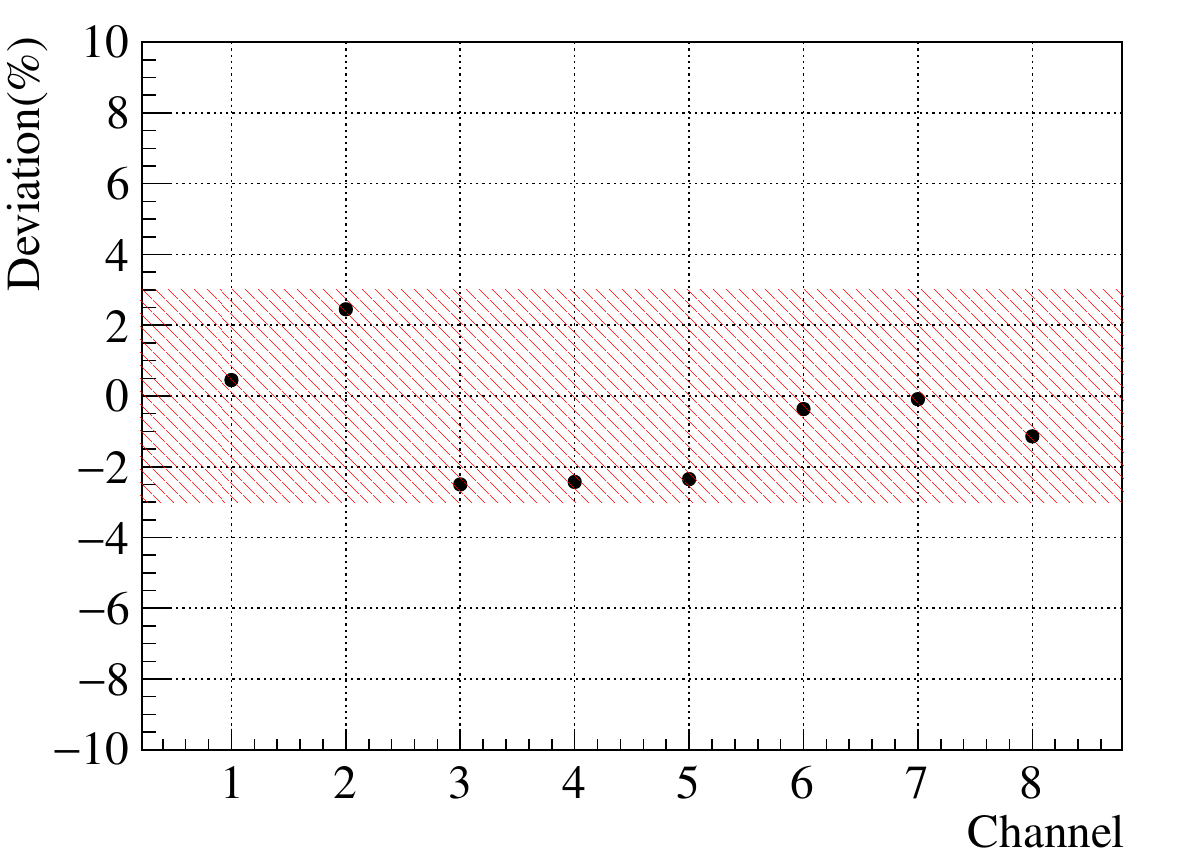}
  \caption{The deviation of 8 channels. The maximum and minimum error band is drawn by backslash.}
  \label{fig:systemError}
\end{figure}

After getting 8 channels' corrected charge ($Q$), the deviation of each channel can be calculated through $Q$ divided by the average of 8 channel's $Q$ (Fig.~\ref{fig:systemError}). The system error is expressed by the difference between the maximum and minimum values, which is within $\pm$ 3\%.

\subsection{The system validation by point-like light source}
\label{sec-pointLight-validation}

\begin{figure}[!hbt]
  \centering
  \includegraphics[width=0.8\textwidth]{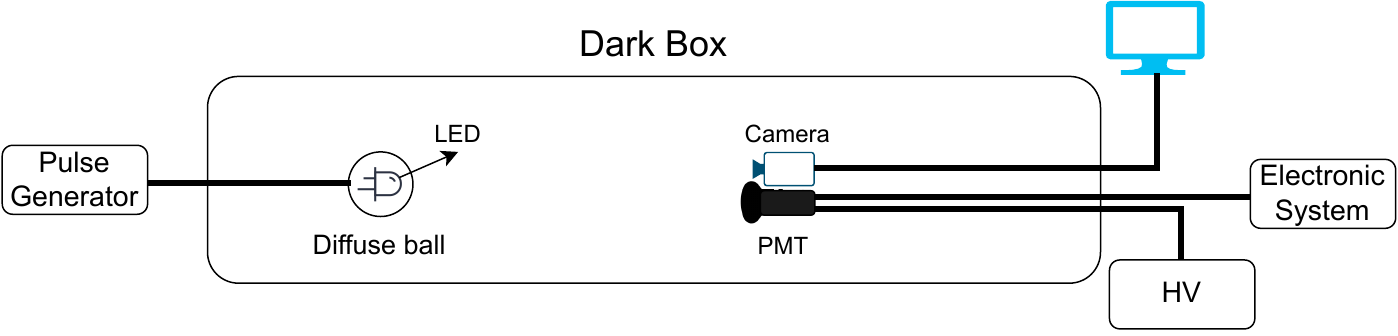}
  \caption{The setup to test the inverse-square law of the point-like source. The PMT and camera are put in the dark box and at different places facing to the diffuser.}
  \label{fig:pmt-camera-r2}
\end{figure}

Because the proposed device is a large-scale design ($\sim$30~m), it is easy to realize in the large Cherenkov detector, however, it is difficult to achieve in the laboratory. But, we can do the principle verification in the laboratory. So, the device in the air with a point light source verification is done. In the future in water, only the medium is different, but the performance of each part of the detector does not depend on the medium. When the light flies in the air, this corresponds to the WAL ($\lambda$) is infinitely long and the equation is simplified into a simpler form:

\begin{equation}
    Q(d)=Ad^{-2},
    \label{eq:inverse-square-law}
\end{equation}
This right is the form of inverse-square law. Similar to the test in the dark box (Fig.~\ref{fig:setup}), an experiment is set up to verify the principle of the proposed device in the air. Both a PMT and a camera are positioned at the same distance from the diffuser (Fig.~\ref{fig:pmt-camera-r2}). The straw-hat LED is enclosed within the Nylon66 diffuse ball. The diffuser was covered with black ABS plastic, leaving only a 5~mm diameter hole facing the PMT and camera. This hole acts as a point-like source. During the test, the PMT and camera were positioned at varying distances to validate the inverse-square law of the light intensity of the system.

\begin{figure}[!hbt]
  \centering
  \includegraphics[width=0.4\textwidth]{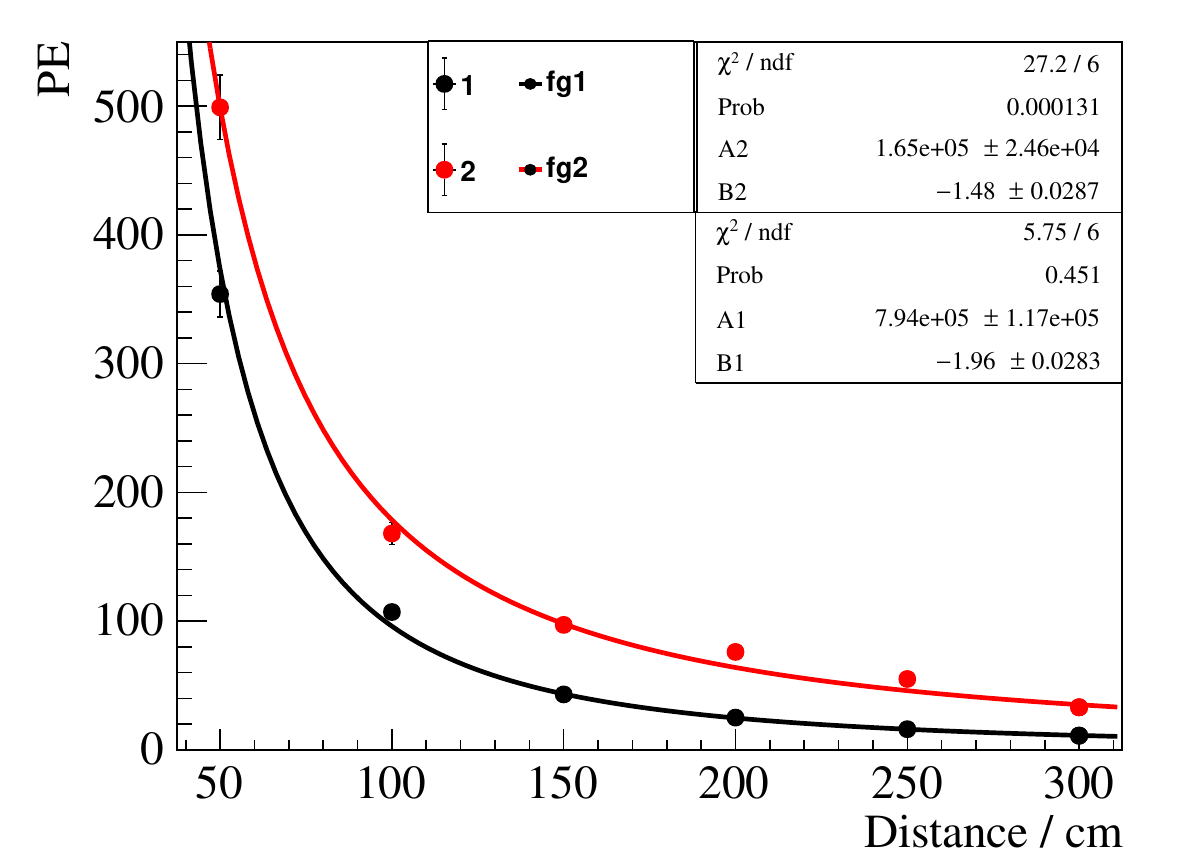}
  \includegraphics[width=0.4\textwidth]{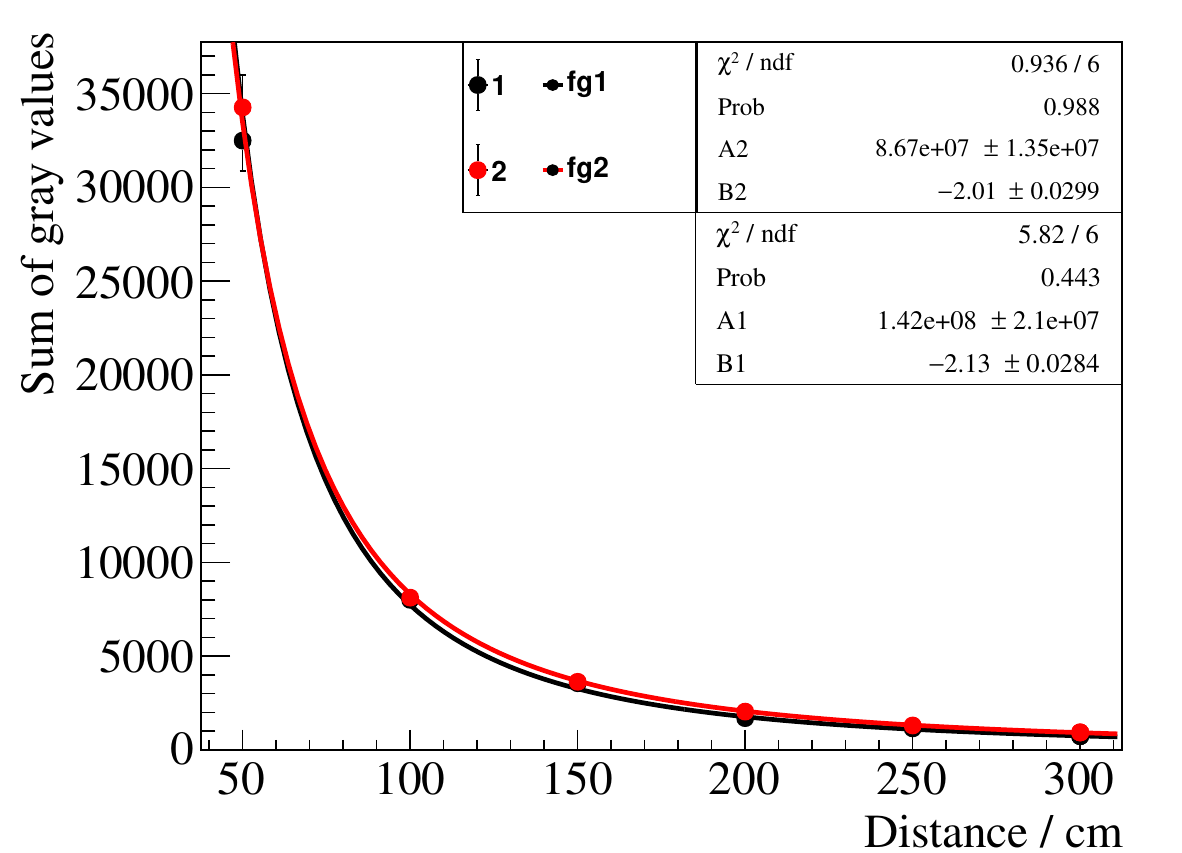}
  \caption{Inverse square law of the point-like light source verification by PMT (left) and camera (right). The red dots and black dots correspond to different experimental configurations of the wall of the dark box, which are HDPE and black cloth, respectively.}
  \label{fig:PMT and camera-r2}
\end{figure}

The PMT and camera are positioned directly facing the light source at six locations, with distances ranging from 50~cm to 300~cm at 50~cm intervals.  (Fig.~\ref{fig:PMT and camera-r2}). Approximately 20,000 events were triggered by a pulse generator for each point, and PMT waveforms were recorded by Flash ADC. The PMT PE number is calculated by integrating the waveforms and dividing by the calibrated gain, resulting in a statistical error of less than 1\%. Considering the system error in section \ref{sec-uncertainty3percent}, the position and angle facing error, a total error of 5\% is set. To verify whether different distance points adhere to the inverse-square law, the fit function is expressed using the exponential equation
\begin{equation}
    Q(d)=Ad^B 
    \label{eq:inverseSquare}
    \end{equation}
Where $Q(d)$ is the photosensor collected charge; $B$ is the exponent, which should be close to -2 for the point-like light source.

The PMT and camera data were taken in the dark box with the wall of black high-density polyethylene (HDPE), corresponding to the red dots in Fig.~\ref{fig:PMT and camera-r2}, while the black dots correspond to the black cloth of dark box. The red and black curves show the exponential fit results by the equation~(\ref{eq:inverseSquare}). Ideally, the value of B should be close to -2, however, B2 from the red curve is -1.48 (Fig.~\ref{fig:PMT and camera-r2}, left), which deviates significantly from -2. The reason is identified as the smooth surface of the black HDPE film reflecting light with mirror reflection at a certain angle. When the black HDPE was replaced by the black cloth material, the mirror reflection disappeared and the fit result of B1 was -1.96, which is close to the ideal case. For the camera measurement (Fig.~\ref{fig:PMT and camera-r2}, right), it can be seen that reflections do not have a significant impact on the camera data, because the camera's imaging principle is by focusing the imaging light and the stray light can not be focused. 

\begin{figure}
  \centering
  \includegraphics[width=0.4\textwidth]{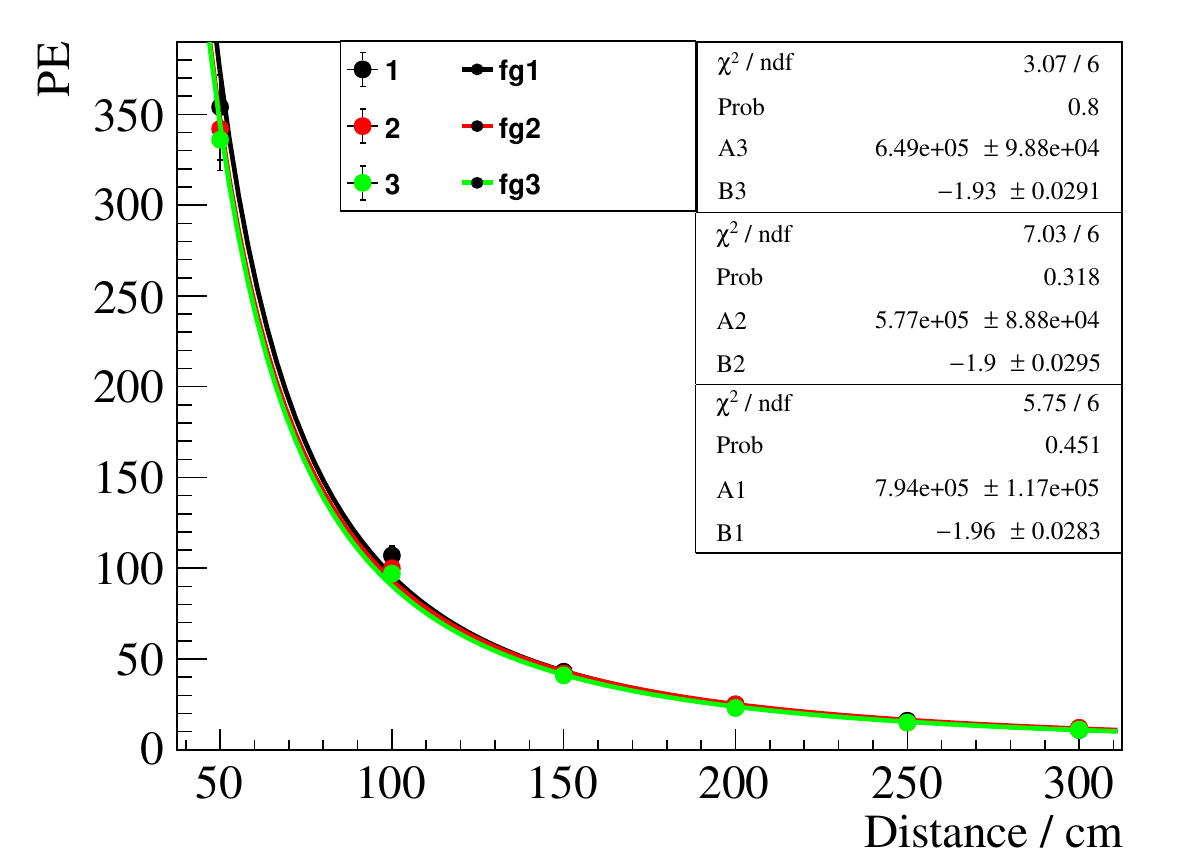}
  \includegraphics[width=0.4\textwidth]{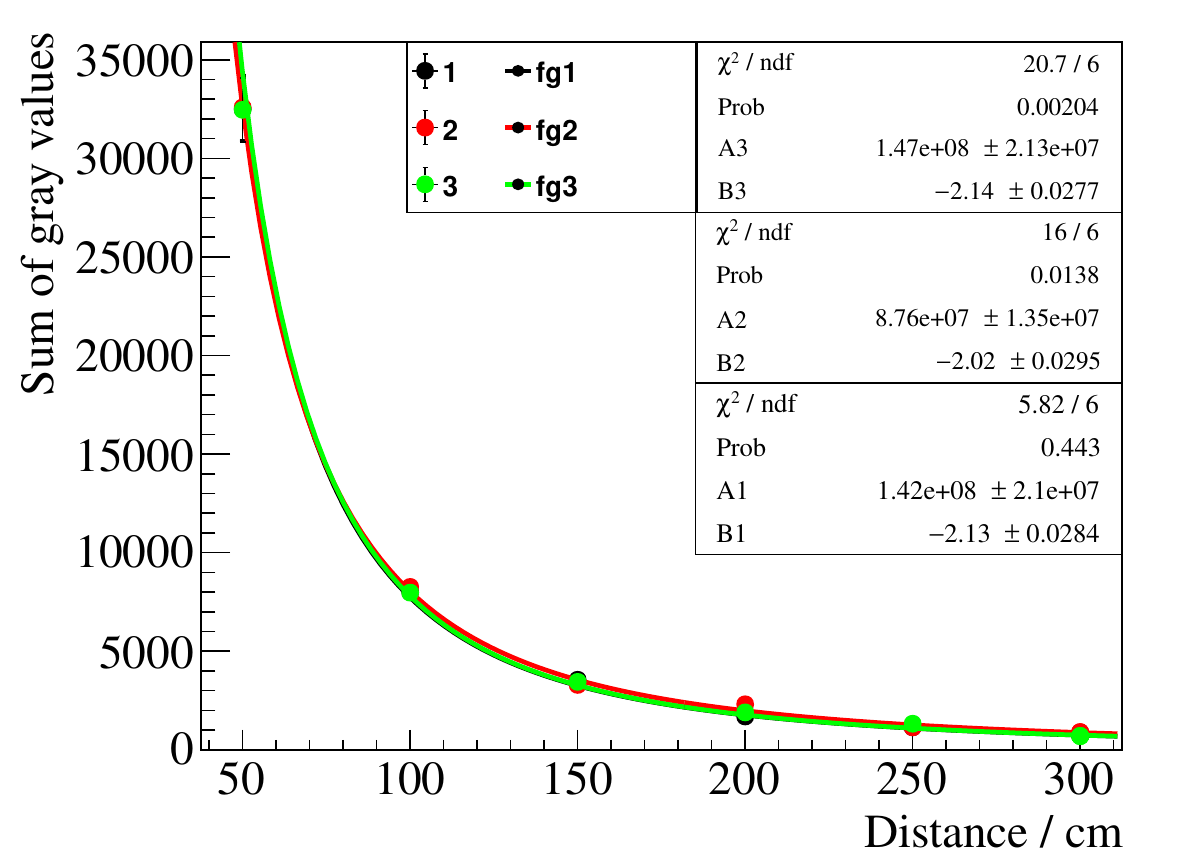}
  \caption{The point-like source test by PMT (left) and camera (right) under the black cloth configuration and the corresponding fit results. The black dots and curves correspond to the photosensors directly facing to the light source; red dots and curves correspond to the photosensor's 3-degree left shift; green dots and curves correspond to the 3-degree right shift.}
  \label{fig:After black camera and PMT-r2}
\end{figure}

While the uniformity of the luminescent spot itself has been validated by camera image, it also requires testing at different angles. In the black cloth configuration, we further studied the different facing angles to verify that the inverse square law is still followed (Fig.~\ref{fig:After black camera and PMT-r2}). For better comparison, the black dots and curve (fg1) are the same as the black-colored data in Fig.~\ref{fig:PMT and camera-r2}. The red and green dots correspond to the PMT and the camera rotated about 3 degrees around the light source for the left and right sides, respectively. Although the different angles, the data are consistent within the error. Both the PMT and the camera can fit well on the $ d^{-2} $ decay curve.

\begin{figure}[!hbt]
  \centering
  \includegraphics[width=0.4\textwidth]{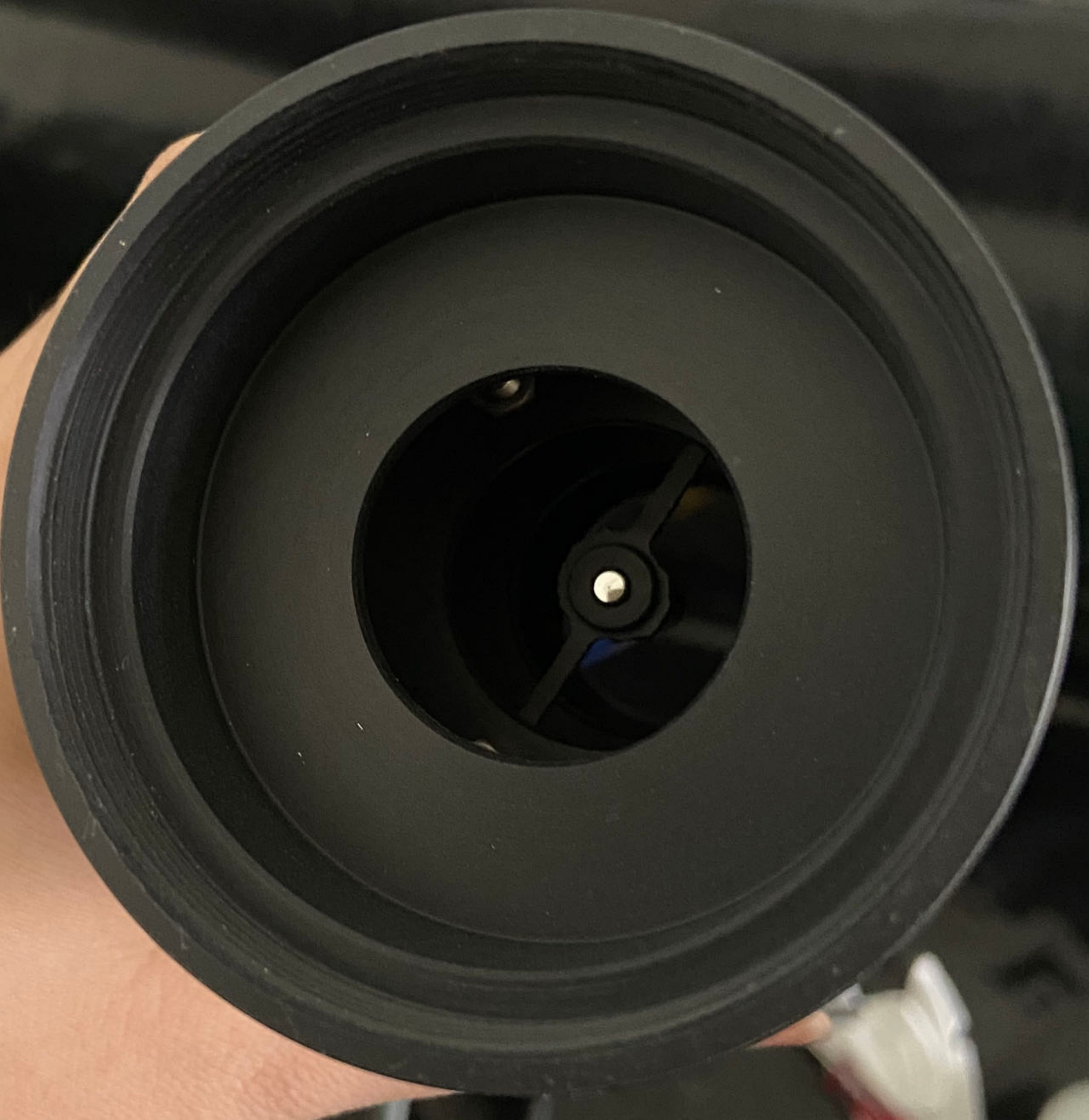}
  \includegraphics[width=0.55\textwidth]{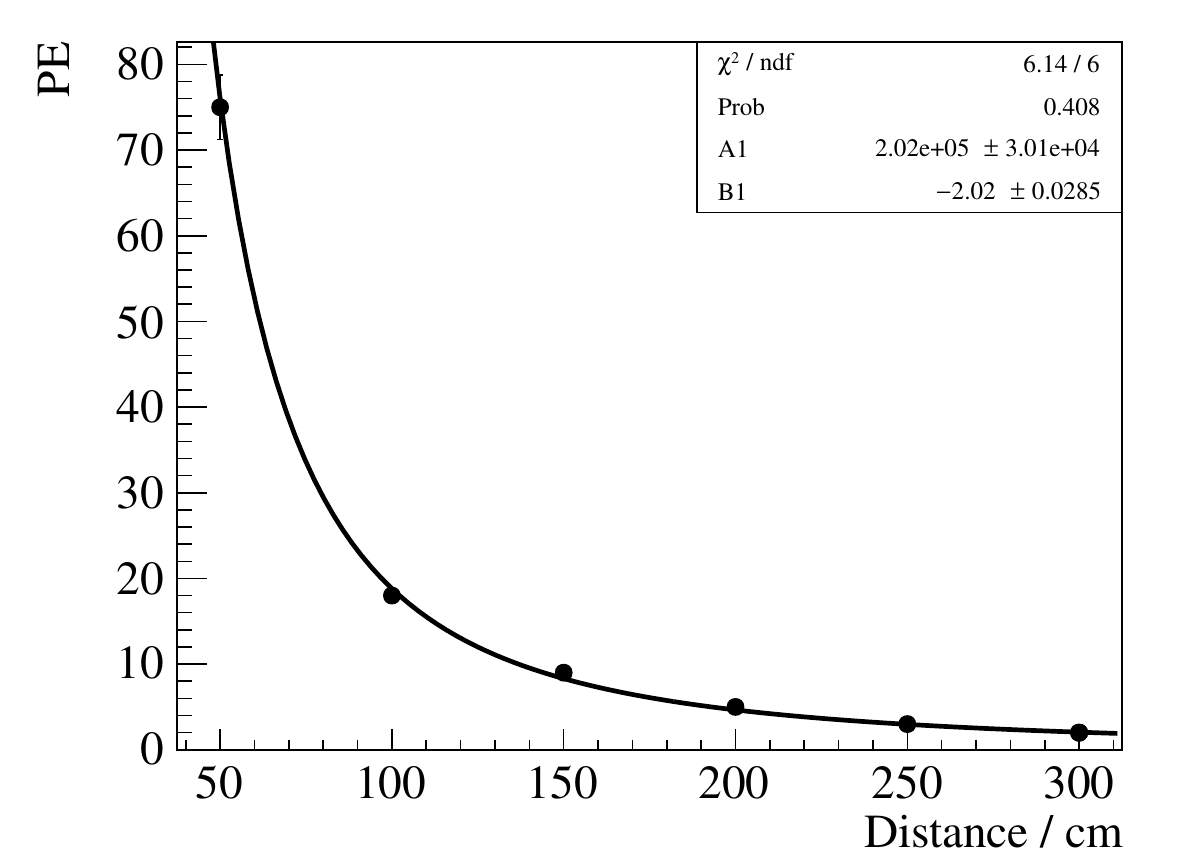}
  \caption{A frontal view of a shutter (left). The point-like source test by PMT using 5 shutters fit results (right).}
  \label{fig:shuttertest}
\end{figure}

To eliminate the influence of reflection and stray light around, shutters were designed to cover the PMT and only let direct light in (Fig.~\ref{fig:shuttertest}, left). Each shutter has a diameter of 7~cm and a length of 3~cm. There is a plate as a light blocker in the middle of the shutter with a hole of diameter 3~cm. Each shutter can be screwed together, in the experiment we found 5 shutters can give the best stray light elimination (Fig.~\ref{fig:shuttertest}, right). The fit parameter B1 is -2.02 which is close to the ideal value -2. The design of the shutter is related to the situation of the detectors used in the future. If some detectors are black inside, it may not be necessary to use a shutter or a complicated design. If the detectors have more reflections in the future, it is necessary to consider a better design for the shutter.

\section{The system performance anticipation}\label{sec-anticipation}

  The most critical aspect of WAL research is controlling overall measurement device errors. Measurement errors include the uniformity error of the LED's diffusing sphere, non-uniformity of the fiber, PMT detection efficiency error, PMT gain calibration error, and non-linearity of the PMT at different light intensities. Section~\ref{sec-uniformityTest} studies showed the non-uniformity error of the LED diffusing sphere is less than 2\%. In the actual measurement process of the PMT, the non-uniformity error of the fiber and the PMT detection efficiency error are coupled together. By setting different measurement conditions, we can separately measure their differences and ultimately obtain the overall system error, which is controlled within a range of 3\% as shown in Fig.~\ref{fig:systemError}. Additionally, relevant measurement results ($<$ 2\%)\cite{LPMTMassProduction} for the error caused by the non-linearity of the PMT at different light intensities ($<$ 600 PEs) are available. Different types of PMTs and the base design give different non-linearity \cite{wudr-nonlinearity}. Therefore, the overall error is estimated to be less than 5\% after calibration and correction.
  
Based on the various errors measured in the experiment, predictions about the entire attenuation length device's performance were made, and simple Monte Carlo (MC) simulations were conducted. Assumed distances from 8 PMTs to the LED source are 3~m, 7~m, 11~m, 15~m, 19~m, 23~m, 27~m, and 31~m, respectively. We assumed the expected measurement capabilities and measurement errors if the device were placed in water with attenuation lengths of 30~m, 40~m, 60~m, 80~m, and 100~m.

  \begin{figure}[!hbt]
  \centering
  \includegraphics[width=0.6\textwidth]{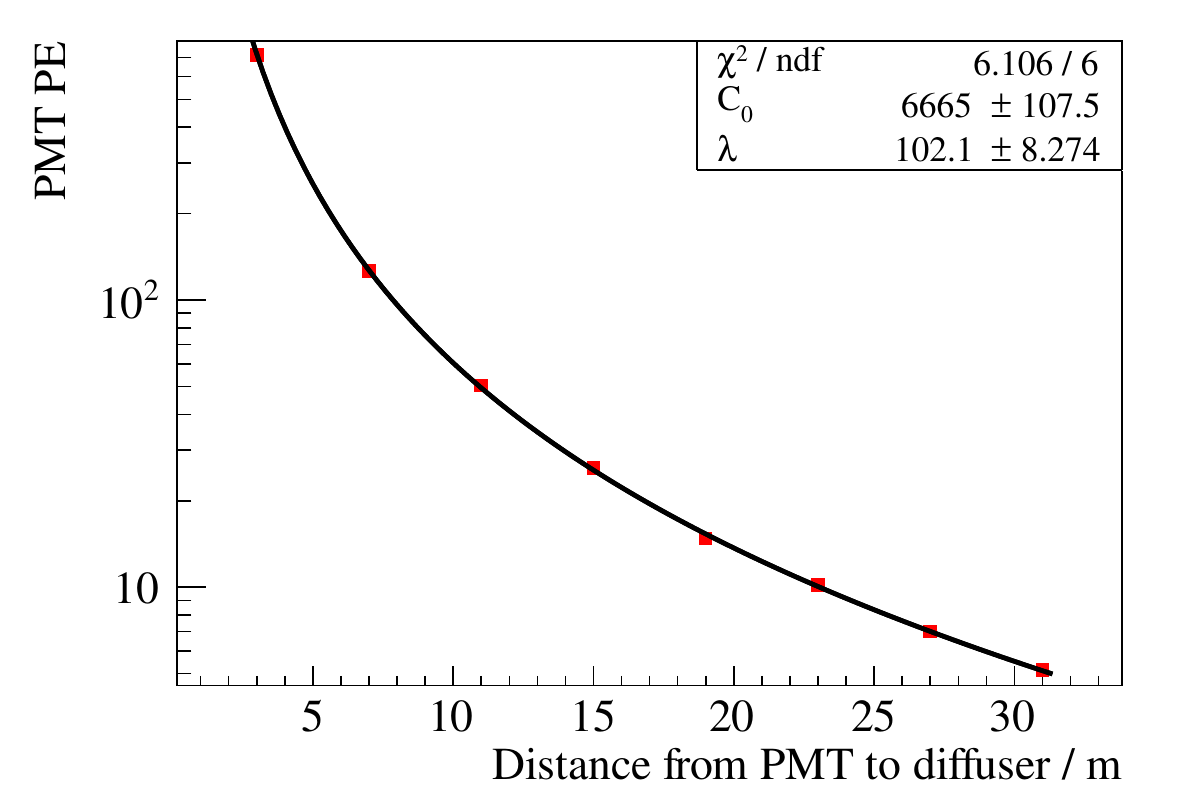}
  \caption{Simulated WAL test result, assumed the water attenuation length is 100~m.}
  \label{fig:simAttL}
\end{figure}

 Following the simulation, the distribution of photoelectrons received by the PMT at each layout position, along with their respective errors can be obtained. By employing exponential decay, the attenuation length curve for each assumed value can be fitted. As Fig.~\ref{fig:simAttL} shown, it's the case for 100~m fitting results. The fitting results reveal varying uncertainties for different attenuation length values listed in Tab.~\ref{tab:attL}. Specifically, if the WAL is 30~m, after fitting the data of 8 PMTs, the uncertainty of the WAL fit result can be controlled to approximately 2.4\%, and for a WAL of 80~m, the test result for WAL is approximately 5.3\%. For the super-long attenuation length, 100~m, it still can be measured with the uncertainty in 8.1\%.
  \begin{table}
	\centering
		\caption{The expected water attenuation length and the uncertainties by toyMC simulation.}
	\label{tab:attL}
	\begin{tabular}{ccc}
		\hline
		\hline
		WAL & Results & Uncertainty \\
		\hline
        30~m & 30.4 $\pm$ 0.7~m & 2.4\%  \\
        40~m & 39.3 $\pm$ 1.2~m & 3.1\%  \\
        60~m &  59.4 $\pm$ 2.4~m & 4.1\%  \\
        80~m &  78.5 $\pm$ 4.2~m & 5.3\%  \\
        100~m &  102.1 $\pm$ 8.3~m & 8.1\%  \\
        \hline
        \hline
	\end{tabular}
	\end{table}

\section{Conclusions}\label{sec7}

A novel water attenuation length measurement device is proposed, which could be directly put in the large Cherenkov detector to give online monitoring without moving operation. The light non-uniformity of the light source and diffuse ball are studied and showed that the overall non-uniformity of the diffuser is less than 2\%. The camera study indicated that the camera should operate in its linear range. A comparative experiment was conducted between the camera and the PMT and the PMT found that the PMT is more susceptible to stray light. Through the prototype study, the system uncertainty can be controlled in 3\% and the verification of the exponential function of the point light source showed a good agreement with a factor of -2 concerning the diffuse reflection of the black wall. If the wall has high reflection, the light shielding should be employed to cover the PMT and avoid the stray light influence. After integrating each part of the detector in our model, the toy MC showed the overall system error is less than 5\%, within the expected error range. The WAL at 40~m can be tested by the device within uncertainty $\sim$ 3\%. Even the WAL super-long at $\sim$ 100~m still can be tested by the device within uncertainty 8.1\%. This project has identified materials that yield better results for measuring the WAL, aiming to improve the accuracy of long WAL measurements. This has significant guiding implications for the WAL measurement on the large scale (> 30 m) of ultrapure Cherenkov detectors, such as JUNO and Hyper-K.

\section{Acknowledgments}

Many thanks for the financial support from the Xie Jialin Foundation of Institute of High Energy Physics (IHEP, No.~E3546EU2), National Natural Science Foundation of China (Grant No.~11975258 and 11875282), the State Key Laboratory of Particle Detection and Electronics (No.~SKLPDE-ZZ-202208).


\begin{thebibliography}{99}


   
   \bibitem{LiuJinchang}
   Liu Jinchang et al., {\it Measurement of Attenuation-Length and Light Yield of Liquid Scintillator}, \href{https://doi.org/10.3321/j.issn:0254-3052.2007.01.017}{HIGH ENERGY PHYSICS AND NUCLEAR PHYSICS, Vol. 31, No. 1, Jan., 2007 (In Chinese)}.
  
      \bibitem{YuBoxiang-PMTAttL}
      GAO Long et al., {\it Attenuation length measurements of a liquid scintillator with LabVIEW and reliability evaluation of the device}, \href{https://doi.org/10.1088/1674-1137/37/7/076001}{Chinese Physics C Vol. 37, No. 7 (2013) 076001}.
      
       \bibitem{HYang-junoLS-JINST}
 H. Yang et al., {\it Light attenuation length of high quality linear alkyl benzene as liquid scintillator solvent for the JUNO experiment}, \href{https://doi.org/10.1088/1748-0221/12/11/T11004}{ 2017 JINST 12 T11004}.
 
        \bibitem{XiangweiYin-junoLS-RDTM}
 Xiang-Wei Yin et al., {\it Precise measurement of attenuation length of the JUNO liquid scintillator}, \href{https://doi.org/10.1007/s41605-020-00185-x}{Radiation Detection Technology and Methods (2020) 4:312–318}. 
      
       \bibitem{Anqi}
       Q. An et al., {\it The performance of a prototype array of water Cherenkov detectors for the LHAASO project}, \href{http://dx.doi.org/10.1016/j.nima.2013.04.081}{Nuclear Instruments and Methods in Physics Research A 724 (2013) 12–19}. 
       
          \bibitem{WCDA}
   Ji Fang et al., {\it Water Quality Monitoring Analysis Based on WCDA}, \href{http://dx.doi.org/10.3969/j.issn.1672-7673.2020.02.015}{Astronomical Research And Technology, Vol. 17, No. 2, Apr. 2020 (In Chinese)}.
   
     \bibitem{chips}
    F. Amat et al., {\it Measuring the attenuation length of water in the CHIPS-M water Cherenkov detector}, \href{http://dx.doi.org/10.1016/j.nima.2016.11.032}{Nuclear Instruments and Methods in Physics Research A 844 (2017) 108–115}.

      
      \bibitem{LHAASO}
     Li, Cong et al., {\it An apparatus to measure water optical attenuation length for LHAASO-MD}, \href{http://doi.org/10.1016/j.nima.2018.03.029}{Nuclear Inst. and Methods in Physics Research, A 892 (2018) 122–126}.
    
    
      \bibitem{Super-KDet}
  S. Fukuda et al., {\it The Super-Kamiokande detector}, \href{http://doi.org/10.1016/S0168-9002(03)00425-X}{Nuclear Instruments and Methods in Physics Research A 501 (2003) 418–462}. 
  
  
\bibitem{Hailing}
Z. P. Ye et al., (TRIDENT Collaboration), {\it Proposal for a neutrino telescope in South China Sea}, \href{
https://doi.org/10.48550/arXiv.2207.04519}{ arXiv:2207.04519v1 [astro-ph.HE], 10 Jul, 2022}.

\bibitem{ANTARES}
J.A. Aguilar et al., (The ANTARES Collaboration), {\it Transmission of light in deep sea water at the site of the ANTARES neutrino telescope}, \href{https://doi.org/10.1016/j.astropartphys.2004.11.006}{Astroparticle Physics 23 (2005) 131–155}. 

      
\bibitem{YuBoxiang-LS}
WANG Xu-kun et al., {\it Development of Automatic Liquid Scintillator's Attenuation Length Measurement System Based on CMOS Camera}, \href{https://doi.org/10.3969/j.issn.0258-0934.2015.12.022}{Nuclear Electronics \& Detection Technology, Vol. 35, No. 12, Dec. 2015, (In Chinese)}.

     \bibitem{CCDfactory}
         \href{http://www.work5power.com/?cate=29}{http://www.work5power.com/?cate=29}
    
    
        \bibitem{CMOS and CCD}
     Durini Daniel, {\it High Performance Silicon Imaging: Fundamentals and Applications of CMOS and CCD Sensors, Second Edition}, \href{https://doi.org/10.1016/C2017-0-01564-1}{Woodhead Publishing, Elsevier Science \& Technology, 2020}. 
     
     
     
    
     \bibitem{CCD photoreceptor area}
     Fan Ming-An., {\it Image Processing Based Detection of Beam Non-parallelism}, Nanjing University Of Information Science \& Technology, 2017 (In Chinese).
    
    
    \bibitem{CCD Excessive saturation}
     Zhang Zhen, Xiang-Ai Cheng, and Zong-Fu Jiang, {\it Excessive saturation effect of visible light CCD}, \href{https://d.wanfangdata.com.cn/periodical/qjgylzs200806009}{Qiang ji guang yu li zi shu 20.6 (2008): 917–920}.
    
        \bibitem{Charge-Coupled Devices}
     Barbe, D. F., {\it Charge-Coupled Devices}, \href{https://doi.org/10.1007/3-540-09832-1}{Topics in Applied Physics, volume 38, Springer-Verlag Berlin Heidelberg, 1980}.
    
    
    \bibitem{Measuring Colour}
     R. W. G. Hunt, and M. R. Pointer. {\it Measuring Colour. Fourth Edition}, Newark: John Wiley \& Sons, Incorporated, 2011. Print.
    
    \bibitem{CCD charge leakage}
     Cai De-Fang, Shi Xiao-Hua, {\it Spectral characterization of CCD detectors with saturation thresholds}, \href{https://doi.org/10.3969/j.issn.0253-2743.2000.03.009}{Ji Guang Za Zhi (Chongqing, China) 21.3 (2000): 14–15, (In Chinese)}. 
    
    
    \bibitem{Reflectance Spectroscopy}
      Gusta, Kort$\ddot{u}$m, {\it Reflectance Spectroscopy: Principles, Methods, Applications}, \href{https://doi.org/10.1007/978-3-642-88071-1}{Springer-Verlag New York Inc., 1969. Web, ISBN: 978-3-642-88073-5}.
      
    

    \bibitem{taobao}
    \href{https://www.taobao.com}{https://www.taobao.com}
    
    \bibitem{3-inch-PMT}
    Chuanya Cao, et al., {\it Mass production and characterization of 3-inch PMTs for the JUNO experiment}, \href{https://doi.org/10.1016/j.nima.2021.165347}{Nuclear Inst. and Methods in Physics Research, A 1005 (2021) 165347}.
    
    \bibitem{fiberFactory}
    \href{http://www.xi-ri.com}{http://www.xi-ri.com}
    
    \bibitem{haiqiong-paper}
    H. Q. Zhang, et al., {\it Comparison on PMT waveform reconstructions with JUNO prototype}, \href{https://doi.org/10.1088/1748-0221/14/08/T08002}{2019 JINST 14 T08002}.

    \bibitem{LPMTMassProduction}
    JUNO collaboration, {\it Mass testing and characterization of 20-inch PMTs for JUNO}, \href{https://doi.org/10.1140/epjc/s10052-022-11002-8}{Eur. Phys. J. C (2022) 82:1168}. 
    
    \bibitem{wudr-nonlinearity}
    Diru Wu et al., {\it Study on the linearity of 20'' dynode and MCP PMTs}, \href{https://doi.org/10.1088/1748-0221/18/05/P05033}{2023 JINST 18 P05033}.


\end{thebibliography}
\end{document}